\documentclass[
aip,apl,%
amsmath,amssymb,
reprint,
]{revtex4-2}
\usepackage{color}
\usepackage{graphicx}
\usepackage{lineno}
\usepackage{gensymb}
\usepackage{textcomp}
\usepackage[dvipdfmx,colorlinks=true]{hyperref}
\hypersetup{
 linkcolor=blue,
 citecolor=blue,
 urlcolor=blue
 }

\begin{document}
\title{Perpendicular magnetic anisotropy in ultra-thin Cu$_2$Sb-type (Mn-Cr)AlGe films fabricated onto thermally oxidized silicon substrates}
\author{Takahide Kubota}
\email[]{takahide.kubota@tohoku.ac.jp}
\thanks{author to whom correspondence should be addressed.}
\affiliation{Institute for Materials Research, Tohoku University, Sendai 980-8577, Japan}
\affiliation{Center for Spintronics Research Network, Tohoku University, Sendai 980-8577, Japan}
\author{Keita Ito}
\affiliation{Institute for Materials Research, Tohoku University, Sendai 980-8577, Japan}
\affiliation{Center for Spintronics Research Network, Tohoku University, Sendai 980-8577, Japan}
\author{Rie Y Umetsu}
\affiliation{Institute for Materials Research, Tohoku University, Sendai 980-8577, Japan}
\affiliation{Center for Spintronics Research Network, Tohoku University, Sendai 980-8577, Japan}
\affiliation{Center for Science and Innovation in Spintronics (Core Research Cluster), Tohoku University, Sendai 980-8577, Japan}
\author{Koki Takanashi}
\affiliation{Institute for Materials Research, Tohoku University, Sendai 980-8577, Japan}
\affiliation{Center for Spintronics Research Network, Tohoku University, Sendai 980-8577, Japan}
\affiliation{Center for Science and Innovation in Spintronics (Core Research Cluster), Tohoku University, Sendai 980-8577, Japan}
%
%
\begin{abstract}
Perpendicularly magnetized films exhibiting small saturation magnetizations ($M_\mathrm{s}$) are essential for spin-transfer-torque magnetoresistive random access memories (STT-MRAMs). In this study, the intermetallic compound {(Mn-Cr)AlGe} with a Cu$_2$Sb crystal structure was investigated as a material exhibiting low $M_\mathrm{s}$ ($\sim 300$ kA/m) and high-perpendicular magnetic anisotropy ($K_\mathrm{u}$). The layer thickness dependence of $K_\mathrm{u}$ and effects of Mg-insertion layers at the top and bottom (Mn-Cr)AlGe$|$MgO interfaces were studied for film samples fabricated onto thermally oxidized silicon substrates to realize high $K_\mathrm{u}$ values in the thickness range of a few nanometers.
The values of $K_\mathrm{u}$ were approximately $7 \times 10^5$ and $2 \times 10^5$ J/m$^3$ at room temperature for 5 and 3 nm-thick (Mn-Cr)AlGe films, respectively, with an optimum annealing temperature of 400 {\degree}C and Mg-insertion thicknesses of 1.4 and 3.0 nm for the top and bottom interfaces, respectively. The Mg insertions were relatively thick compared with results of similar studies of the insertion effect on magnetic tunnel junctions. Cross-sectional transmission electron microscope images revealed that the Mg-insertion layers acted as barriers to the interdiffusion of Al atoms as well as oxidization from the MgO layers. The $K_\mathrm{u}$ at a few-nanometer thicknesses was comparable to or higher than those reported for perpendicularly magnetized CoFeB films, which are conventionally used in MRAMs, whereas the $M_\mathrm{s}$ value was one third or smaller than those of the CoFeB films.
The developed (Mn-Cr)AlGe films are promising materials because of their magnetic properties and their compatibility to the silicon process in film fabrication.
\end{abstract}
\date{\today}%
%
\maketitle
%
Magnetoresistive random access memory (MRAM) is an emerging memory that has attracted significant interest because of its non-volatility and fast operation.~\cite{Bhatti2017,Ikegawa2020,Na2021}
To expand the practical applications of MRAMs, the critical current density ($J_\mathrm{c}$) must be reduced for the magnetization switching induced by the spin-transfer torque (STT) in magnetic tunnel junctions (MTJs). The  $J_\mathrm{c}$ is proportional to saturation magnetization ($M_\mathrm{s}$) of the magnetization switching layer\cite{Slonczewski1996}; thus, magnetic materials with low $M_\mathrm{s}$ values are important. In addition, for STT-MRAMs with gigabit-class memory capacities, perpendicular magnetization is essential for small $J_\mathrm{c}$~ values\cite{Kishi2008,Yoda2010} as well as for high-density memory cells. Moreover, a high perpendicular magnetic anisotropy ($K_\mathrm{u}$) is necessary for data retention.~\cite{Bhatti2017,Ikegawa2020,Na2021}
Various materials have been studied to determine a class of low $M_\mathrm{s}$ and high $K_\mathrm{u}$ systems, \textit{e.g.}, D0$_{22}$-type Mn$_3$Ga alloys,~\cite{Wu2009,Mizukami2011,Mizukami2012} Mn$_3$Ge alloys,~\cite{Kurt2012,Mizukami2013,Sugihara2014}, and body-centered-tetragonal Mn ultra-thin films.\cite{Suzuki2018} Although attractive properties were reported for all these materials exhibiting $M_\mathrm{s}$ in the range $\sim$100--300 kA/m and a $K_\mathrm{u}$ of $\sim$10$^6$ J/m$^3$ or greater at room temperature, most of the studies on these materials were conducted using single crystalline MgO substrates, which are unsuitable for practical devices. A few studies reported on stacking film samples onto thermally oxidized silicon substrates, \textit{e.g.} using Mn--Ga alloys;~\cite{Ono2017,Suzuki2017} however, another technical problem is that all the Mn alloys used in previous studies were ferrimagnets, which decrease the spin polarization at the interface cyanowing to experimentally unavoidable atomic steps.\cite{Jeong2016}

This study focused on an intermetallic compound (Mn-Cr)AlGe~\cite{Kubota2019} as an alternative material to the Mn alloys used in previous studies. The (Mn-Cr)AlGe compound is a ferromagnet that belongs to a family of Mn-based intermetallic compounds exhibiting the Cu$_2$Sb crystal structure, such as MnAlGe~\cite{Shibata1973,Umetsu2014,Mizukami2013a,Kubota2019,Kubota2020,Umetsu2021a} and MnGaGe.~\cite{Goodenough1975,Umetsu2014,Sun2020}
The $M_\mathrm{s}$ values of Mn-compounds range from 200 to 300 kA/m, which are relatively small because the small magnetic moment of the ferromagnetically coupled Mn-moments has an itinerant electronic structure.~\cite{Motizuki2010}
The values of $K_\mathrm{u}$ at room temperature are 7.3 $\times$ 10$^5$ and 8.1 $\times$ 10$^5$ J/m$^3$ for (Mn$_{0.76}$Cr$_{0.28}$)Al$_{0.94}$Ge$_{1.02}$\cite{Kubota2019} and Mn$_{1.0}$Ga$_{1.1}$Ge$_{0.9}$\cite{Sun2020}, respectively, which are significantly higher than those reported for conventionally used CoFeB films with perpendicular magnetization.~\cite{Ikeda2010} These physical properties of low $M_\mathrm{s}$ and high $K_\mathrm{u}$ are promising for MRAM applications. Another merit is the silicon process compatibility of these compounds, \textit{i.e.}, films with (001) textures exhibiting perpendicular magnetization are available on the SiO$_2$ surfaces as well as other amorphous substrates.~\cite{Sawatzky1973,Sherwood1971,Lee1973,Kubota2019,Kubota2020,Sun2020t}
The (001) orientation is also an essential requirement for MTJs using the (001)-textured MgO$|$CoFe(B) structure that exhibits large tunnel magnetoresistance.~\cite{Yuasa2007} 
In addition, a recent study simulated the reduction of $J_\mathrm{c}$ in an antiferromagnetically coupled layered structure consisting of high- and low-$M_\mathrm{s}$ materials,~\cite{Yamada2020} for which the (Mn-Cr)AlGe can be a low-$M_\mathrm{s}$ material layered with, \textit{e.g.}, CoFe(B) exhibiting a relatively high $M_\mathrm{s}$.
 In our previous study, the template effect was demonstrated for the (001)-textured growth of MnAlGe films using (001)-textured MgO seed/capping layers, in which the MgO(001) template promotes the (001) texture in the MnAlGe films; however, a relatively thick dead layer around the MnAlGe$|$MgO interfaces with the degraded $K_\mathrm{u}$ for a thickness less than 10 nm.~\cite{Kubota2021}
The sputtered MgO layer may cause oxidization at the interface with a metallic magnetic material, which can be eliminated by inserting an ultra-thin metallic layer, \textit{e.g.}, a Mg layer, in between.~\cite{Tsunekawa2005,Sakuraba2007,Kubota2011mg}
Such an interface layer possibly improves the dead layer problem of the interface between the Mn compounds and MgO; thus, in this study, the effect of the Mg-interface layer was investigated for (Mn-Cr)AlGe$|$MgO interfaces to realize a high $K_\mathrm{u}$ with a layer thickness of a few nanometers.

\begin{figure}
\includegraphics[clip,scale=0.9]{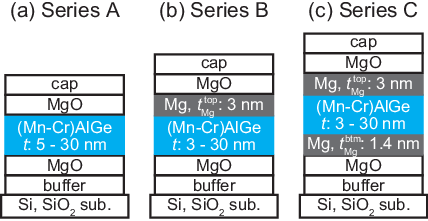}
\caption{
Stacking structures of the sample series (a) A, (b) B, and (c) C.
}
\label{fig:series}
\end{figure}

%
The layered film samples were fabricated using a magnetron sputtering system with a base vacuum pressure below 1 $\times$ 10$^{-7}$ Pa. All the samples were fabricated onto single crystalline silicon (100) substrates with a 650 nm-thick thermally oxidized amorphous layer on the surface. The stacking structure of the samples was as follows: Sub.$|$buffer layers$|$MgO 1.5 nm$|$Mg $t_\mathrm{Mg}^\mathrm{btm}|$(Mn-Cr)AlGe $t|$Mg $t_\mathrm{Mg}^\mathrm{top}|$MgO 1.5 nm$|$cap, where $t_\mathrm{Mg}^\mathrm{btm (top)}$ and $t$ are the thicknesses of the Mg insertion for the bottom (top) interface and (Mn-Cr)AlGe layer, respectively.
The buffer layers were Ta 3 nm$|$W 0.3 nm$|$CoFeBTa 1 nm, from bottom to top, which partly assumed the bottom electrodes of MTJs. Note that the 0.3 nm-thick W layer was not used in the series A samples; however, no difference in magnetic properties was confirmed by comparing the samples with and without the W layer.
Note that, in this study, the CoFeBTa layer was nonmagnetic to clearly study the magnetic properties of the (Mn-Cr)AlGe layer.
The film composition of the (Mn-Cr)AlGe layer was (Mn$_{0.8}$Cr$_{0.2}$)Al$_{1.1}$Ge$_{0.9}$ by atomic ratio. 
All the layers were deposited at room temperature inside the ultra-high vacuum chamber, and a 3 nm-thick Ta was deposited on the top as a cap layer. After the capping, all samples were annealed using a vacuum furnace. The annealing temperatures were 300, 400, and 500 {\textdegree}C.
Magnetization curves were measured at room temperature using vibration sample magnetometers and a superconducting quantum interference device magnetometer. The interface properties and crystal structures were characterized using high-angle annular dark field (HAADF)-scanning transmission electron microscope (STEM) images.
The thicknesses of the Mg insertion were optimized via two steps: First, the top interface was optimized by changing $t_\mathrm{Mg}^\mathrm{top}$ from 0 to 7 nm, for which $t_\mathrm{Mg}^\mathrm{btm}$ was fixed to 0. Subsequently, for the second step, the bottom interface was optimized with $t_\mathrm{Mg}^\mathrm{btm}$ in the range 0--1.4 nm, for which $t_\mathrm{Mg}^\mathrm{top}$ was fixed to 3.0 nm. For both steps, $t$ was fixed to 10 nm. After the two steps, three sample series were selected as shown in Fig.~\ref{fig:series} to study the $t$ dependence of $K_\mathrm{u}$ for the (Mn-Cr)AlGe layer, for which $t$ was changed from 3 to 30 nm.
%
\begin{figure}
\includegraphics[clip,scale=1.0]{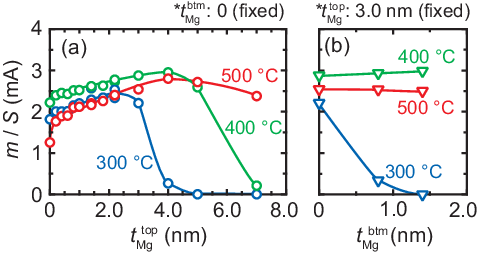}
\caption{Mg-insertion thickness ($t_\mathrm{Mg}^\mathrm{top}$ or $t_\mathrm{Mg}^\mathrm{btm}$) dependence of magnetization values ($m$) per the sample area ($S$). $t = 10$ nm for all samples. (a) In the $t_\mathrm{Mg}^\mathrm{top}$ dependence, $t_\mathrm{Mg}^\mathrm{btm}$ was fixed to 0, and (b) in the bottom $t_\mathrm{Mg}^\mathrm{btm}$ dependence, $t_\mathrm{Mg}^\mathrm{top}$ was fixed to 3.0 nm. The temperatures shown in the panels are the annealing temperatures of the samples.}
\label{arealM}
\end{figure}
The Mg-insertion dependence of magnetization ($m$) per sample area, ($S$) is shown in Fig. \ref{arealM}. For the top interface, shown in Fig. \ref{arealM}(a), the magnetization increased with $t_\mathrm{Mg}^\mathrm{top}$ up to the certain thickness values depending on the annealing temperature, \textit{i.e.}, the values of $t_\mathrm{Mg}^\mathrm{top}$ for the maximum $m/S$ shifted toward larger values for higher annealing temperature, the reason for which will be discussed in a latter part of this article.
The maximum $m/S$ was approximately 2.9 mA for $t_\mathrm{Mg}^\mathrm{top}$ = 3--4 nm and the annealing temperature was 400 {\textdegree}C; thus, a $t_\mathrm{Mg}^\mathrm{top}$ of 3.0~nm was selected as an optimum value for the first step.
The Mg insertion for the bottom interface, shown in Fig.~\ref{arealM}(b), exhibited moderate dependence except at the annealing temperature of 300~{\textdegree}C, at which $m/S$ monotonically decreased with $t_\mathrm{Mg}^\mathrm{btm}$. For other annealing temperatures, with increasing $t_\mathrm{Mg}^\mathrm{btm}$, $m/S$ gradually increased at 400~{\textdegree}C, while it gradually decreased at 500 {\textdegree}C. Based on the second step, the $t_\mathrm{Mg}^\mathrm{btm}$ of 1.4 nm was selected.
%
%
%
The three sample series listed in Fig.~\ref{fig:series} were selected to investigate the $t$ dependence of $K_\mathrm{u}$.
\begin{figure}
\includegraphics[clip,scale=1.0]{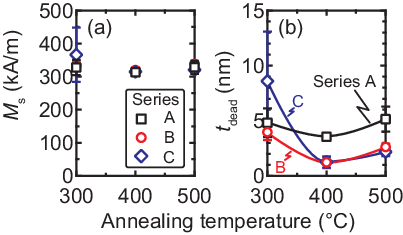}
\caption{
Annealing temperature dependence of (a) saturation magnetization ($M_\mathrm{s}$) and the thicknesses of dead layer ($t_\mathrm{dead}$) evaluated using the $t$ dependence of $m/S$ plots.
}
\label{Ms}
\end{figure}
The $M_\mathrm{s}$ and the dead-layer thickness ($t_\mathrm{dead}$) were evaluated using the slope and the intercept to the horizontal axis, respectively, in the plots of $m/S$ as a function of $t$. The values are summarized in Fig.~\ref{Ms}. The $M_\mathrm{s}$ exhibited no dependence on the sample series or annealing temperature within the range of error bars. The values were approximately 330 kA/m, which was close to the theoretical value of 352 kA/m for (Mn$_{0.8}$Cr$_{0.2}$)AlGe.~\cite{Kubota2019} In contrast, $t_\mathrm{dead}$ exhibited clear dependence on the sample series and annealing temperature: for the annealing temperature dependence, $t_\mathrm{dead}$ had minima at 400 {\textdegree}C for all sample series. Among the sample series, series B and C had a minima, a $t_\mathrm{dead}$ of approximately 1.2 $\pm$ 0.5 nm, which was smaller than that of 3.5 $\pm$ 0.2 nm for series A. 
\begin{figure}
\includegraphics[clip,scale=1.0]{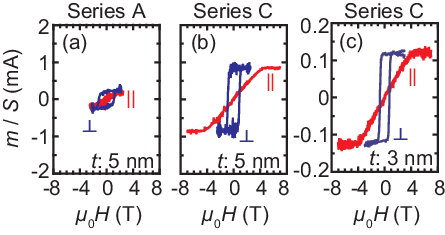}
\caption{
Magnetization curves of 5 nm-thick samples for (a) series A and (b) C. (c) A 3 nm-thick sample for series C. The annealing temperature was 400 {\textdegree}C for all samples. The marks $\perp$ and $||$ represent applied magnetic fields in the perpendicular-to-plane and in-plane directions, respectively. 
}
\label{MH}
\end{figure}
The differences among the sample series were clearly observed in the magnetic properties, particularly at $t$ below 10 nm. Fig. \ref{MH} shows the magnetization curves of the 5 nm-thick samples for series A and C, and a 3 nm-thick sample for series C.
The squareness of hysteresis curves for the perpendicular-to-plane magnetic field direction was approximately 1 for both 5 nm-thick samples of series A and C; however, the magnitude of magnetization for the sample in series A (Fig.~\ref{MH}(a))was smaller than that in series C (Fig.~\ref{MH}(b)). In series C, the perpendicular magnetization with a high squareness ratio was maintained down to a $t$ of 3 nm (Fig.~\ref{MH}(c)).
\begin{figure}
\includegraphics[clip,scale=1.0]{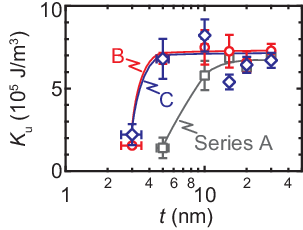}
\caption{
$t$ dependence of the perpendicular magnetic anisotropy energy ($K_\mathrm{u}$.)}
\label{Ku}
\end{figure}
The $t$ dependence of $K_\mathrm{u}$ was evaluated to quantitatively characterize the differences among the sample series. $K_\mathrm{u}$ was defined using the actual layer thickness )$t - t_\mathrm{dead}$) with the following equation: $K_\mathrm{u} = K_\mathrm{u}^\mathrm{eff}/(t - t_\mathrm{dead}) + 
(\mu_0/2)M_\mathrm{s}^2$, where $K_\mathrm{u}^\mathrm{eff}$ is the effective perpendicular magnetic anisotropy energy per unit area. This was calculated using the area enclosed by the perpendicular-to-plane and in-plane magnetization curves, e.g., as shown in Fig.~\ref{MH}. The $K_\mathrm{u}$ values for all the sample series are summarized in Fig.~\ref{Ku} as a function of $t$. For series A, a $K_\mathrm{u}$ of approximately $7 \times 10^5$ J/m$^3$ was achieved in samples down to a $t$ of 15 nm, and at a $t$ of 10 nm or thinner, $K_\mathrm{u}$ decreased with decreasing $t$, which was approximately $1 \times 10^5$ J/m$^3$ at 5 nm. By contrast, for series B and C, a $K_\mathrm{u}$ of approximately $7 \times 10^5$ J/m$^3$ was maintained down to a $t$ of 5 nm; in addition, 3 nm-thick samples had a $K_\mathrm{u}$ of approximately $2 \times 10^5$ J/m$^3$. These results clearly confirmed that the reduction of $t_\mathrm{dead}$ efficiently improved the $K_\mathrm{u}$ in the few-nanometer-thickness region.
\begin{figure*}
\includegraphics[clip,scale=0.8]{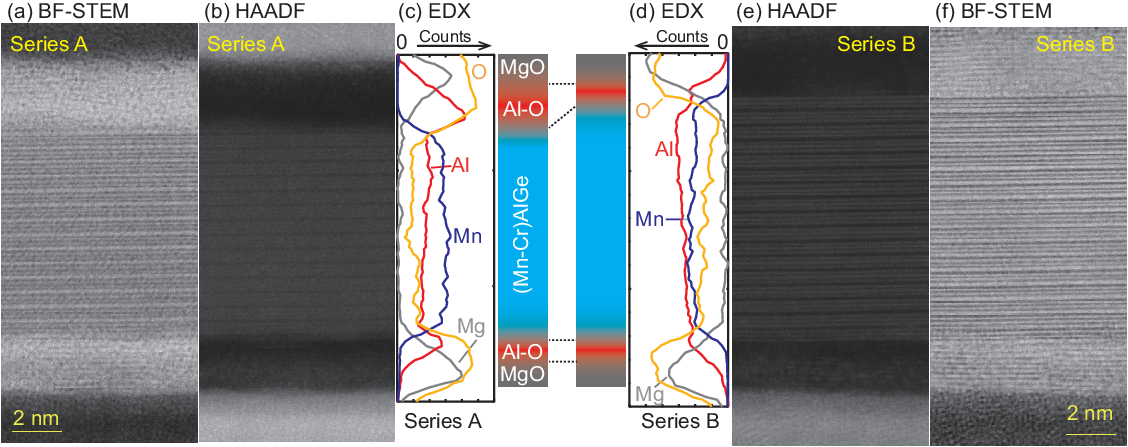}
\caption{
Cross-sectional STEM images of samples for (a, b) series A and (e, f) B. The annealing temperature was 400 {\textdegree}C and $t = 10$ nm for both samples. (a, f) Bright field (BF) images. (b, e) HAADF images. The line profiles of the energy dispersive x-ray (EDX) mapping are indicated in the middle for samples (c) A and (d) B, respectively, with color bars describing the possible layer components.
}
\label{TEM}
\end{figure*}
Although the Mg insertion was observed to be useful for suppressing the $t_\mathrm{dead}$, a challenge that remains is the relatively thick $t_\mathrm{Mg}^\mathrm{top(btm)}$, \textit{i.e.}, in previous studies using MgO-MTJs, the optimum thicknesses of Mg-insertion ranged from 0.4 to 1 nm.\cite{Tsunekawa2005,Sakuraba2007,Kubota2011mg} In this study, the optimum thicknesses were 1.4 and 3 nm for the bottom and top (Mn-Cr)AlGe$|$MgO interfaces, respectively, which were thicker than those in the previous studies.
Cross-sectional HAADF-STEM observations were conducted to characterize the dead layer and crystal structures more clearly, and the images are shown in Fig.~\ref{TEM} for $t = 10$ nm in series A and B.
First, for the crystal structure of the middle region, both samples exhibited high-(001) orientations, as shown in the lattice images of the bright field (BF)-STEM images in Figs.~\ref{TEM}(a) and \ref{TEM}(f) for series A and B, respectively. The high orientation can also be confirmed in the HAADF images shown in Figs.~\ref{TEM}(b) and \ref{TEM}(e), in which periodical contrast changes are clearly indicated. The HAADF contrasts also implied the layered atomic components with the chemical order of the Cu$_2$Sb-type crystal structure.~\cite{Kubota2019,Shibata1973,Umetsu2014,Mizukami2013a,Goodenough1975,Sun2020}
Second, a difference was observed for the actual layer thickness of the (Mn-Cr)AlGe layer: From the BF-STEM images, the layer thicknesses were 8.5 $\pm$ 0.2 and 9.7 $\pm$ 0.2 nm for series A and B, respectively.
Line profiles of energy dispersive X-ray (EDX) spectroscopy indicated the difference in the upper interfaces more clearly, which caused the difference in the layer thicknesses: For the series A sample, an Al-rich oxidized region was observed between the (Mn-Cr)AlGe and top MgO layer, for which the thickness was approximately 2 nm; this is marked as ''Al--O'' in the color scale beside the HAADF image of Fig.~\ref{TEM}(b). By contrast, for series B, a similar Al-rich region was observed at the top interface; however, its thickness of less than 1 nm was significantly smaller than that in series A.
For the interface at the bottom MgO layer, the Al--O layer was also observed in each sample series, for which a thickness of approximately 1 nm was similarly observed.
Such formation of the Al--O layer caused by the interdiffusion of Al atoms from metallic to oxide layer has been reported for layered samples consisting of an ultra-thin Co$_2$FeAl film capped by a sputtered MgO layer. Here, the diffusion of Al atoms might occur during the deposition of the MgO upper layer onto the metallic layer containing Al.~\cite{Wen2016CFA,Husain2019}
The formation of the Al--O layer in this study can be qualitatively understood using the following scenario:
The (Mn-Cr)AlGe layer is amorphous or nano-crystalline in the as-deposited state, for which the constituent  elements are randomly distributed at the surface before the deposition of the MgO layer. During the deposition of the MgO layer, the surface of (Mn-Cr)AlGe could be exposed to a small amount of oxygen plasma, which results in an oxygen-rich interface between the (Mn-Cr)AlGe and the top MgO layers. Here, the free energy of formation of the Al-rich oxide is the smallest among the constituent elements of (Mn-Cr)AlGe~\cite{Ellingham1944}; thus, Al diffusion may occur with the assistance of plasma energy during the deposition of MgO as well as the post-annealing procedure, resulting in the formation of the Al-rich oxide region for the upper interface. For the bottom interface, Al diffusion may also occur during the annealing process; however, the amount of Al-diffusion is less than at the top interface because the oxygen-rich interface is not formed for the bottom interface.
Thus, the Mg insertion is considered to suppress such Al diffusion and the formation of the Al--O layer, resulting in the suppression of $t_\mathrm{dead}$ and the enhancement of $K_\mathrm{u}$ at $t$ in the few-nanometer-thickness region.
Based on the aforementioned discussion, the rationale for the different optimum $t_\mathrm{Mg}^\mathrm{top(btm)}$ values depending on the annealing temperatures shown in Fig.~\ref{arealM} can be obtained: By considering that the diffusion process of Al promotes the crystallization of (Mn-Cr)AlGe for the no Mg-insertion sample, Al atoms move inside the layer with the assistance of a relatively low energy, \textit{i.e.}, the low annealing temperature of 300 {\textdegree}C, which is sufficient for crystallization. However, as Mg insertion increases, the Al diffusion is suppressed, which requires an annealing temperature of 400 {\textdegree}C or higher for crystallization with large magnetizations.
For future device applications in MTJs, it should be noted that the relatively thick Mg-insertion layers cause large tunnel resistance, which is unfavorable. However, the thick Mg-insertion layers are considered to be required, mostly because of the oxygen-plasma damage during the deposition process of the MgO layer. The damage can be controlled by maintaining the machine, \textit{e.g.}, replacing the magnetron cathode for sputtering with another one equipped with a stronger magnet, or adopting a different fabrication process of MgO, such as natural oxidation of Mg.~\cite{Seki2010} With further optimization in engineering, the perpendicularly magnetized ultra-thin (Mn-Cr)AlGe can be applied to MgO MTJs in practical devices.

In summary, the perpendicular magnetic anisotropy of Cu$_2$Sb-type (Mn-Cr)AlGe films was investigated with particular focus on the a-few-nanometer-thickness region of the (Mn-Cr)AlGe layer. Mg-insertion layers were employed at the (Mn-Cr)AlGe$|$MgO interfaces, for which the effects on the magnetic properties and $t_\mathrm{dead}$ were systematically examined.
The $M_\mathrm{s}$ exhibited almost no dependence on the Mg-insertion thickness, whereas $t_\mathrm{dead}$ decreased with the Mg-insertion. Cross-sectional HAADF-STEM images indicated a clear contrast change for the (Mn-Cr)AlGe layer, which represented highly oriented (001)-crystal planes and the chemical order of (Mn-Cr)AlGe. The STEM images also revealed the function of the Mg-insertion as a diffusion barrier to Al atoms.
A $K_\mathrm{u}$ of approximately $7 \times 10^5$ J/m$^3$ at a $t$ of 5 nm is comparable to those of thick samples, \textit{e.g.} 30-nm samples, and much higher than those reported in perpendicularly magnetized CoFeB films, which are conventionally used in STT-MRAMs. Although a $K_\mathrm{u}$ of approximately $2 \times 10^5$ J/m$^3$ at a $t$ of 3 nm is comparable with those of CoFeB films, the small $M_\mathrm{s}$ of (Mn-Cr)AlGe can facilitate reducing the $J_\mathrm{c}$.
Among various materials exhibiting small $M_\mathrm{s}$ and high $K_\mathrm{u}$ values, another merit of (Mn-Cr)AlGe is the silicon process compatibility: All the samples in this study were fabricated using thermally oxidized silicon substrates with relatively thin underlayers, including the (001)-oriented MgO template.\cite{Kubota2021} The solid-phase-epitaxy based method for crystallization is compatible with device fabrication processes for complementary metal oxide semiconductor devices as well as for MTJs.\cite{Yuasa2007}
The (Mn-Cr)AlGe film exhibiting perpendicular magnetization is also interesting from the application perspective.
\section*{Supplementary Material}
Following results are available in Supplementary Material:
All magnetization curves for Figs.~\ref{arealM}  and \ref{Ms},
$t$ dependence of $m/S$ for all samples,
plots of $K_\mathrm{u}$ as a function of the actual layer thickness ($t - t_\mathrm{dead}$),
EDX line profiles for all elements of series A and B,
and an XRD profile of a 30-nm-thick sample in series A in the as-deposited state.

\begin{acknowledgments}
TK would like to thank Dr.~Yoshiaki Sonobe for a fruitful discussion on the topic, Mr.~Issei Narita for technical support of film composition analysis, and the Foundation for Promotion of Material Science and Technology of Japan for HAADF-STEM observation and analysis.
This work was partially supported by KAKENHI (JP20K05296) from JSPS, and a cooperative research program (No.~20G0414) of the CRDAM-IMR, Tohoku University.
\end{acknowledgments}
\section*{Data Availability Statements}
The data that support the findings of this study are available within its supplementary material.
\bibliographystyle{aipnum4-1}
\bibliography{
d:/Lab/Manuscripts/BibTex/MRAM,
d:/Lab/Manuscripts/BibTex/Mn-alloys,
d:/Lab/Manuscripts/BibTex/Heusler_etc
}

\begin{thebibliography}{40}%
\makeatletter
\providecommand \@ifxundefined [1]{%
 \@ifx{#1\undefined}
}%
\providecommand \@ifnum [1]{%
 \ifnum #1\expandafter \@firstoftwo
 \else \expandafter \@secondoftwo
 \fi
}%
\providecommand \@ifx [1]{%
 \ifx #1\expandafter \@firstoftwo
 \else \expandafter \@secondoftwo
 \fi
}%
\providecommand \natexlab [1]{#1}%
\providecommand \enquote  [1]{``#1''}%
\providecommand \bibnamefont  [1]{#1}%
\providecommand \bibfnamefont [1]{#1}%
\providecommand \citenamefont [1]{#1}%
\providecommand \href@noop [0]{\@secondoftwo}%
\providecommand \href [0]{\begingroup \@sanitize@url \@href}%
\providecommand \@href[1]{\@@startlink{#1}\@@href}%
\providecommand \@@href[1]{\endgroup#1\@@endlink}%
\providecommand \@sanitize@url [0]{\catcode `\\12\catcode `\$12\catcode
  `\&12\catcode `\#12\catcode `\^12\catcode `\_12\catcode `\%12\relax}%
\providecommand \@@startlink[1]{}%
\providecommand \@@endlink[0]{}%
\providecommand \url  [0]{\begingroup\@sanitize@url \@url }%
\providecommand \@url [1]{\endgroup\@href {#1}{\urlprefix }}%
\providecommand \urlprefix  [0]{URL }%
\providecommand \Eprint [0]{\href }%
\providecommand \doibase [0]{http://dx.doi.org/}%
\providecommand \selectlanguage [0]{\@gobble}%
\providecommand \bibinfo  [0]{\@secondoftwo}%
\providecommand \bibfield  [0]{\@secondoftwo}%
\providecommand \translation [1]{[#1]}%
\providecommand \BibitemOpen [0]{}%
\providecommand \bibitemStop [0]{}%
\providecommand \bibitemNoStop [0]{.\EOS\space}%
\providecommand \EOS [0]{\spacefactor3000\relax}%
\providecommand \BibitemShut  [1]{\csname bibitem#1\endcsname}%
\let\auto@bib@innerbib\@empty
\bibitem [{\citenamefont {Bhatti}\ \emph {et~al.}(2017)\citenamefont {Bhatti},
  \citenamefont {Sbiaa}, \citenamefont {Hirohata}, \citenamefont {Ohno},
  \citenamefont {Fukami},\ and\ \citenamefont {Piramanayagam}}]{Bhatti2017}%
  \BibitemOpen
  \bibfield  {author} {\bibinfo {author} {\bibfnamefont {S.}~\bibnamefont
  {Bhatti}}, \bibinfo {author} {\bibfnamefont {R.}~\bibnamefont {Sbiaa}},
  \bibinfo {author} {\bibfnamefont {A.}~\bibnamefont {Hirohata}}, \bibinfo
  {author} {\bibfnamefont {H.}~\bibnamefont {Ohno}}, \bibinfo {author}
  {\bibfnamefont {S.}~\bibnamefont {Fukami}}, \ and\ \bibinfo {author}
  {\bibfnamefont {S.}~\bibnamefont {Piramanayagam}},\ }\href {\doibase
  10.1016/j.mattod.2017.07.007} {\bibfield  {journal} {\bibinfo  {journal}
  {Mater Today}\ }\textbf {\bibinfo {volume} {20}},\ \bibinfo {pages} {530}
  (\bibinfo {year} {2017})}\BibitemShut {NoStop}%
\bibitem [{\citenamefont {Ikegawa}\ \emph {et~al.}(2020)\citenamefont
  {Ikegawa}, \citenamefont {Mancoff}, \citenamefont {Janesky},\ and\
  \citenamefont {Aggarwal}}]{Ikegawa2020}%
  \BibitemOpen
  \bibfield  {author} {\bibinfo {author} {\bibfnamefont {S.}~\bibnamefont
  {Ikegawa}}, \bibinfo {author} {\bibfnamefont {F.~B.}\ \bibnamefont
  {Mancoff}}, \bibinfo {author} {\bibfnamefont {J.}~\bibnamefont {Janesky}}, \
  and\ \bibinfo {author} {\bibfnamefont {S.}~\bibnamefont {Aggarwal}},\ }\href
  {\doibase 10.1109/ted.2020.2965403} {\bibfield  {journal} {\bibinfo
  {journal} {{IEEE} Transactions on Electron Devices}\ }\textbf {\bibinfo
  {volume} {67}},\ \bibinfo {pages} {1407} (\bibinfo {year}
  {2020})}\BibitemShut {NoStop}%
\bibitem [{\citenamefont {Na}, \citenamefont {Kang},\ and\ \citenamefont
  {Jung}(2021)}]{Na2021}%
  \BibitemOpen
  \bibfield  {author} {\bibinfo {author} {\bibfnamefont {T.}~\bibnamefont
  {Na}}, \bibinfo {author} {\bibfnamefont {S.~H.}\ \bibnamefont {Kang}}, \ and\
  \bibinfo {author} {\bibfnamefont {S.-O.}\ \bibnamefont {Jung}},\ }\href
  {\doibase 10.1109/tcsii.2020.3040425} {\bibfield  {journal} {\bibinfo
  {journal} {{IEEE} Transactions on Circuits and Systems {II}: Express Briefs}\
  }\textbf {\bibinfo {volume} {68}},\ \bibinfo {pages} {12} (\bibinfo {year}
  {2021})}\BibitemShut {NoStop}%
\bibitem [{\citenamefont {Slonczewski}(1996)}]{Slonczewski1996}%
  \BibitemOpen
  \bibfield  {author} {\bibinfo {author} {\bibfnamefont {J.}~\bibnamefont
  {Slonczewski}},\ }\href {\doibase 10.1016/0304-8853(96)00062-5} {\bibfield
  {journal} {\bibinfo  {journal} {J. Magn. Magn. Mater.}\ }\textbf {\bibinfo
  {volume} {159}},\ \bibinfo {pages} {L1} (\bibinfo {year} {1996})}\BibitemShut
  {NoStop}%
\bibitem [{\citenamefont {Kishi}\ \emph {et~al.}(2008)\citenamefont {Kishi},
  \citenamefont {Yoda}, \citenamefont {Kai}, \citenamefont {Nagase},
  \citenamefont {Kitagawa}, \citenamefont {Yoshikawa}, \citenamefont
  {Nishiyama}, \citenamefont {Daibou}, \citenamefont {Nagamine}, \citenamefont
  {Amano}, \citenamefont {Takahashi}, \citenamefont {Nakayama}, \citenamefont
  {Shimomura}, \citenamefont {Aikawa}, \citenamefont {Ikegawa}, \citenamefont
  {Yuasa}, \citenamefont {Yakushiji}, \citenamefont {Kubota}, \citenamefont
  {Fukushima}, \citenamefont {Oogane}, \citenamefont {Miyazaki},\ and\
  \citenamefont {Ando}}]{Kishi2008}%
  \BibitemOpen
  \bibfield  {author} {\bibinfo {author} {\bibfnamefont {T.}~\bibnamefont
  {Kishi}}, \bibinfo {author} {\bibfnamefont {H.}~\bibnamefont {Yoda}},
  \bibinfo {author} {\bibfnamefont {T.}~\bibnamefont {Kai}}, \bibinfo {author}
  {\bibfnamefont {T.}~\bibnamefont {Nagase}}, \bibinfo {author} {\bibfnamefont
  {E.}~\bibnamefont {Kitagawa}}, \bibinfo {author} {\bibfnamefont
  {M.}~\bibnamefont {Yoshikawa}}, \bibinfo {author} {\bibfnamefont
  {K.}~\bibnamefont {Nishiyama}}, \bibinfo {author} {\bibfnamefont
  {T.}~\bibnamefont {Daibou}}, \bibinfo {author} {\bibfnamefont
  {M.}~\bibnamefont {Nagamine}}, \bibinfo {author} {\bibfnamefont
  {M.}~\bibnamefont {Amano}}, \bibinfo {author} {\bibfnamefont
  {S.}~\bibnamefont {Takahashi}}, \bibinfo {author} {\bibfnamefont
  {M.}~\bibnamefont {Nakayama}}, \bibinfo {author} {\bibfnamefont
  {N.}~\bibnamefont {Shimomura}}, \bibinfo {author} {\bibfnamefont
  {H.}~\bibnamefont {Aikawa}}, \bibinfo {author} {\bibfnamefont
  {S.}~\bibnamefont {Ikegawa}}, \bibinfo {author} {\bibfnamefont
  {S.}~\bibnamefont {Yuasa}}, \bibinfo {author} {\bibfnamefont
  {K.}~\bibnamefont {Yakushiji}}, \bibinfo {author} {\bibfnamefont
  {H.}~\bibnamefont {Kubota}}, \bibinfo {author} {\bibfnamefont
  {A.}~\bibnamefont {Fukushima}}, \bibinfo {author} {\bibfnamefont
  {M.}~\bibnamefont {Oogane}}, \bibinfo {author} {\bibfnamefont
  {T.}~\bibnamefont {Miyazaki}}, \ and\ \bibinfo {author} {\bibfnamefont
  {K.}~\bibnamefont {Ando}},\ }in\ \href {\doibase 10.1109/iedm.2008.4796680}
  {\emph {\bibinfo {booktitle} {2008 {IEEE} International Electron Devices
  Meeting}}}\ (\bibinfo  {publisher} {{IEEE}},\ \bibinfo {year}
  {2008})\BibitemShut {NoStop}%
\bibitem [{\citenamefont {Yoda}\ \emph {et~al.}(2010)\citenamefont {Yoda},
  \citenamefont {Kishi}, \citenamefont {Nagase}, \citenamefont {Yoshikawa},
  \citenamefont {Nishiyama}, \citenamefont {Kitagawa}, \citenamefont {Daibou},
  \citenamefont {Amano}, \citenamefont {Shimomura}, \citenamefont {Takahashi},
  \citenamefont {Kai}, \citenamefont {Nakayama}, \citenamefont {Aikawa},
  \citenamefont {Ikegawa}, \citenamefont {Nagamine}, \citenamefont {Ozeki},
  \citenamefont {Mizukami}, \citenamefont {Oogane}, \citenamefont {Ando},
  \citenamefont {Yuasa}, \citenamefont {Yakushiji}, \citenamefont {Kubota},
  \citenamefont {Suzuki}, \citenamefont {Nakatani}, \citenamefont {Miyazaki},\
  and\ \citenamefont {Ando}}]{Yoda2010}%
  \BibitemOpen
  \bibfield  {author} {\bibinfo {author} {\bibfnamefont {H.}~\bibnamefont
  {Yoda}}, \bibinfo {author} {\bibfnamefont {T.}~\bibnamefont {Kishi}},
  \bibinfo {author} {\bibfnamefont {T.}~\bibnamefont {Nagase}}, \bibinfo
  {author} {\bibfnamefont {M.}~\bibnamefont {Yoshikawa}}, \bibinfo {author}
  {\bibfnamefont {K.}~\bibnamefont {Nishiyama}}, \bibinfo {author}
  {\bibfnamefont {E.}~\bibnamefont {Kitagawa}}, \bibinfo {author}
  {\bibfnamefont {T.}~\bibnamefont {Daibou}}, \bibinfo {author} {\bibfnamefont
  {M.}~\bibnamefont {Amano}}, \bibinfo {author} {\bibfnamefont
  {N.}~\bibnamefont {Shimomura}}, \bibinfo {author} {\bibfnamefont
  {S.}~\bibnamefont {Takahashi}}, \bibinfo {author} {\bibfnamefont
  {T.}~\bibnamefont {Kai}}, \bibinfo {author} {\bibfnamefont {M.}~\bibnamefont
  {Nakayama}}, \bibinfo {author} {\bibfnamefont {H.}~\bibnamefont {Aikawa}},
  \bibinfo {author} {\bibfnamefont {S.}~\bibnamefont {Ikegawa}}, \bibinfo
  {author} {\bibfnamefont {M.}~\bibnamefont {Nagamine}}, \bibinfo {author}
  {\bibfnamefont {J.}~\bibnamefont {Ozeki}}, \bibinfo {author} {\bibfnamefont
  {S.}~\bibnamefont {Mizukami}}, \bibinfo {author} {\bibfnamefont
  {M.}~\bibnamefont {Oogane}}, \bibinfo {author} {\bibfnamefont
  {Y.}~\bibnamefont {Ando}}, \bibinfo {author} {\bibfnamefont {S.}~\bibnamefont
  {Yuasa}}, \bibinfo {author} {\bibfnamefont {K.}~\bibnamefont {Yakushiji}},
  \bibinfo {author} {\bibfnamefont {H.}~\bibnamefont {Kubota}}, \bibinfo
  {author} {\bibfnamefont {Y.}~\bibnamefont {Suzuki}}, \bibinfo {author}
  {\bibfnamefont {Y.}~\bibnamefont {Nakatani}}, \bibinfo {author}
  {\bibfnamefont {T.}~\bibnamefont {Miyazaki}}, \ and\ \bibinfo {author}
  {\bibfnamefont {K.}~\bibnamefont {Ando}},\ }\href {\doibase
  10.1016/j.cap.2009.12.021} {\bibfield  {journal} {\bibinfo  {journal} {Curr.
  Appl Phys.}\ }\textbf {\bibinfo {volume} {10}},\ \bibinfo {pages} {e87}
  (\bibinfo {year} {2010})}\BibitemShut {NoStop}%
\bibitem [{\citenamefont {Wu}\ \emph {et~al.}(2009)\citenamefont {Wu},
  \citenamefont {Mizukami}, \citenamefont {Watanabe}, \citenamefont {Naganuma},
  \citenamefont {Oogane}, \citenamefont {Ando},\ and\ \citenamefont
  {Miyazaki}}]{Wu2009}%
  \BibitemOpen
  \bibfield  {author} {\bibinfo {author} {\bibfnamefont {F.}~\bibnamefont
  {Wu}}, \bibinfo {author} {\bibfnamefont {S.}~\bibnamefont {Mizukami}},
  \bibinfo {author} {\bibfnamefont {D.}~\bibnamefont {Watanabe}}, \bibinfo
  {author} {\bibfnamefont {H.}~\bibnamefont {Naganuma}}, \bibinfo {author}
  {\bibfnamefont {M.}~\bibnamefont {Oogane}}, \bibinfo {author} {\bibfnamefont
  {Y.}~\bibnamefont {Ando}}, \ and\ \bibinfo {author} {\bibfnamefont
  {T.}~\bibnamefont {Miyazaki}},\ }\href {\doibase 10.1063/1.3108085}
  {\bibfield  {journal} {\bibinfo  {journal} {Appl. Phys. Lett.}\ }\textbf
  {\bibinfo {volume} {94}},\ \bibinfo {pages} {122503} (\bibinfo {year}
  {2009})}\BibitemShut {NoStop}%
\bibitem [{\citenamefont {Mizukami}\ \emph {et~al.}(2011)\citenamefont
  {Mizukami}, \citenamefont {Wu}, \citenamefont {Sakuma}, \citenamefont
  {Walowski}, \citenamefont {Watanabe}, \citenamefont {Kubota}, \citenamefont
  {Zhang}, \citenamefont {Naganuma}, \citenamefont {Oogane}, \citenamefont
  {Ando},\ and\ \citenamefont {Miyazaki}}]{Mizukami2011}%
  \BibitemOpen
  \bibfield  {author} {\bibinfo {author} {\bibfnamefont {S.}~\bibnamefont
  {Mizukami}}, \bibinfo {author} {\bibfnamefont {F.}~\bibnamefont {Wu}},
  \bibinfo {author} {\bibfnamefont {A.}~\bibnamefont {Sakuma}}, \bibinfo
  {author} {\bibfnamefont {J.}~\bibnamefont {Walowski}}, \bibinfo {author}
  {\bibfnamefont {D.}~\bibnamefont {Watanabe}}, \bibinfo {author}
  {\bibfnamefont {T.}~\bibnamefont {Kubota}}, \bibinfo {author} {\bibfnamefont
  {X.}~\bibnamefont {Zhang}}, \bibinfo {author} {\bibfnamefont
  {H.}~\bibnamefont {Naganuma}}, \bibinfo {author} {\bibfnamefont
  {M.}~\bibnamefont {Oogane}}, \bibinfo {author} {\bibfnamefont
  {Y.}~\bibnamefont {Ando}}, \ and\ \bibinfo {author} {\bibfnamefont
  {T.}~\bibnamefont {Miyazaki}},\ }\href {\doibase
  10.1103/physrevlett.106.117201} {\bibfield  {journal} {\bibinfo  {journal}
  {Phys. Rev. Lett.}\ }\textbf {\bibinfo {volume} {106}},\ \bibinfo {pages}
  {117201} (\bibinfo {year} {2011})}\BibitemShut {NoStop}%
\bibitem [{\citenamefont {Mizukami}\ \emph {et~al.}(2012)\citenamefont
  {Mizukami}, \citenamefont {Kubota}, \citenamefont {Wu}, \citenamefont
  {Zhang}, \citenamefont {Miyazaki}, \citenamefont {Naganuma}, \citenamefont
  {Oogane}, \citenamefont {Sakuma},\ and\ \citenamefont {Ando}}]{Mizukami2012}%
  \BibitemOpen
  \bibfield  {author} {\bibinfo {author} {\bibfnamefont {S.}~\bibnamefont
  {Mizukami}}, \bibinfo {author} {\bibfnamefont {T.}~\bibnamefont {Kubota}},
  \bibinfo {author} {\bibfnamefont {F.}~\bibnamefont {Wu}}, \bibinfo {author}
  {\bibfnamefont {X.}~\bibnamefont {Zhang}}, \bibinfo {author} {\bibfnamefont
  {T.}~\bibnamefont {Miyazaki}}, \bibinfo {author} {\bibfnamefont
  {H.}~\bibnamefont {Naganuma}}, \bibinfo {author} {\bibfnamefont
  {M.}~\bibnamefont {Oogane}}, \bibinfo {author} {\bibfnamefont
  {A.}~\bibnamefont {Sakuma}}, \ and\ \bibinfo {author} {\bibfnamefont
  {Y.}~\bibnamefont {Ando}},\ }\href {\doibase 10.1103/physrevb.85.014416}
  {\bibfield  {journal} {\bibinfo  {journal} {Phys. Rev. B}\ }\textbf {\bibinfo
  {volume} {85}},\ \bibinfo {pages} {014416} (\bibinfo {year}
  {2012})}\BibitemShut {NoStop}%
\bibitem [{\citenamefont {Kurt}\ \emph {et~al.}(2012)\citenamefont {Kurt},
  \citenamefont {Baadji}, \citenamefont {Rode}, \citenamefont {Venkatesan},
  \citenamefont {Stamenov}, \citenamefont {Sanvito},\ and\ \citenamefont
  {Coey}}]{Kurt2012}%
  \BibitemOpen
  \bibfield  {author} {\bibinfo {author} {\bibfnamefont {H.}~\bibnamefont
  {Kurt}}, \bibinfo {author} {\bibfnamefont {N.}~\bibnamefont {Baadji}},
  \bibinfo {author} {\bibfnamefont {K.}~\bibnamefont {Rode}}, \bibinfo {author}
  {\bibfnamefont {M.}~\bibnamefont {Venkatesan}}, \bibinfo {author}
  {\bibfnamefont {P.}~\bibnamefont {Stamenov}}, \bibinfo {author}
  {\bibfnamefont {S.}~\bibnamefont {Sanvito}}, \ and\ \bibinfo {author}
  {\bibfnamefont {J.~M.~D.}\ \bibnamefont {Coey}},\ }\href {\doibase
  10.1063/1.4754123} {\bibfield  {journal} {\bibinfo  {journal} {Appl. Phys.
  Lett.}\ }\textbf {\bibinfo {volume} {101}},\ \bibinfo {pages} {132410}
  (\bibinfo {year} {2012})}\BibitemShut {NoStop}%
\bibitem [{\citenamefont {Mizukami}\ \emph
  {et~al.}(2013{\natexlab{a}})\citenamefont {Mizukami}, \citenamefont {Sakuma},
  \citenamefont {Sugihara}, \citenamefont {Kubota}, \citenamefont {Kondo},
  \citenamefont {Tsuchiura},\ and\ \citenamefont {Miyazaki}}]{Mizukami2013}%
  \BibitemOpen
  \bibfield  {author} {\bibinfo {author} {\bibfnamefont {S.}~\bibnamefont
  {Mizukami}}, \bibinfo {author} {\bibfnamefont {A.}~\bibnamefont {Sakuma}},
  \bibinfo {author} {\bibfnamefont {A.}~\bibnamefont {Sugihara}}, \bibinfo
  {author} {\bibfnamefont {T.}~\bibnamefont {Kubota}}, \bibinfo {author}
  {\bibfnamefont {Y.}~\bibnamefont {Kondo}}, \bibinfo {author} {\bibfnamefont
  {H.}~\bibnamefont {Tsuchiura}}, \ and\ \bibinfo {author} {\bibfnamefont
  {T.}~\bibnamefont {Miyazaki}},\ }\href {\doibase 10.7567/APEX.6.123002}
  {\bibfield  {journal} {\bibinfo  {journal} {Appl. Phys Express}\ }\textbf
  {\bibinfo {volume} {6}},\ \bibinfo {pages} {123002} (\bibinfo {year}
  {2013}{\natexlab{a}})}\BibitemShut {NoStop}%
\bibitem [{\citenamefont {Sugihara}\ \emph {et~al.}(2014)\citenamefont
  {Sugihara}, \citenamefont {Mizukami}, \citenamefont {Yamada}, \citenamefont
  {Koike},\ and\ \citenamefont {Miyazaki}}]{Sugihara2014}%
  \BibitemOpen
  \bibfield  {author} {\bibinfo {author} {\bibfnamefont {A.}~\bibnamefont
  {Sugihara}}, \bibinfo {author} {\bibfnamefont {S.}~\bibnamefont {Mizukami}},
  \bibinfo {author} {\bibfnamefont {Y.}~\bibnamefont {Yamada}}, \bibinfo
  {author} {\bibfnamefont {K.}~\bibnamefont {Koike}}, \ and\ \bibinfo {author}
  {\bibfnamefont {T.}~\bibnamefont {Miyazaki}},\ }\href {\doibase
  10.1063/1.4870625} {\bibfield  {journal} {\bibinfo  {journal} {Appl. Phys.
  Lett.}\ }\textbf {\bibinfo {volume} {104}},\ \bibinfo {pages} {132404}
  (\bibinfo {year} {2014})}\BibitemShut {NoStop}%
\bibitem [{\citenamefont {Suzuki}\ \emph {et~al.}(2018)\citenamefont {Suzuki},
  \citenamefont {Kimura}, \citenamefont {Kubota},\ and\ \citenamefont
  {Mizukami}}]{Suzuki2018}%
  \BibitemOpen
  \bibfield  {author} {\bibinfo {author} {\bibfnamefont {K.~Z.}\ \bibnamefont
  {Suzuki}}, \bibinfo {author} {\bibfnamefont {S.}~\bibnamefont {Kimura}},
  \bibinfo {author} {\bibfnamefont {H.}~\bibnamefont {Kubota}}, \ and\ \bibinfo
  {author} {\bibfnamefont {S.}~\bibnamefont {Mizukami}},\ }\href {\doibase
  10.1021/acsami.8b15606} {\bibfield  {journal} {\bibinfo  {journal} {{ACS}
  Appl. Mater. {\&} Interfaces}\ }\textbf {\bibinfo {volume} {10}},\ \bibinfo
  {pages} {43305} (\bibinfo {year} {2018})}\BibitemShut {NoStop}%
\bibitem [{\citenamefont {Ono}\ \emph {et~al.}(2017)\citenamefont {Ono},
  \citenamefont {Suzuki}, \citenamefont {Ranjbar}, \citenamefont {Sugihara},\
  and\ \citenamefont {Mizukami}}]{Ono2017}%
  \BibitemOpen
  \bibfield  {author} {\bibinfo {author} {\bibfnamefont {A.}~\bibnamefont
  {Ono}}, \bibinfo {author} {\bibfnamefont {K.~Z.}\ \bibnamefont {Suzuki}},
  \bibinfo {author} {\bibfnamefont {R.}~\bibnamefont {Ranjbar}}, \bibinfo
  {author} {\bibfnamefont {A.}~\bibnamefont {Sugihara}}, \ and\ \bibinfo
  {author} {\bibfnamefont {S.}~\bibnamefont {Mizukami}},\ }\href {\doibase
  10.7567/apex.10.023005} {\bibfield  {journal} {\bibinfo  {journal} {Appl.
  Phys. Express}\ }\textbf {\bibinfo {volume} {10}},\ \bibinfo {pages} {023005}
  (\bibinfo {year} {2017})}\BibitemShut {NoStop}%
\bibitem [{\citenamefont {Suzuki}\ \emph {et~al.}(2017)\citenamefont {Suzuki},
  \citenamefont {Ono}, \citenamefont {Ranjbar}, \citenamefont {Sugihara},\ and\
  \citenamefont {Mizukami}}]{Suzuki2017}%
  \BibitemOpen
  \bibfield  {author} {\bibinfo {author} {\bibfnamefont {K.~Z.}\ \bibnamefont
  {Suzuki}}, \bibinfo {author} {\bibfnamefont {A.}~\bibnamefont {Ono}},
  \bibinfo {author} {\bibfnamefont {R.}~\bibnamefont {Ranjbar}}, \bibinfo
  {author} {\bibfnamefont {A.}~\bibnamefont {Sugihara}}, \ and\ \bibinfo
  {author} {\bibfnamefont {S.}~\bibnamefont {Mizukami}},\ }\href {\doibase
  10.1109/tmag.2017.2700494} {\bibfield  {journal} {\bibinfo  {journal} {{IEEE}
  Trans. Magn.}\ }\textbf {\bibinfo {volume} {53}},\ \bibinfo {pages} {2101004}
  (\bibinfo {year} {2017})}\BibitemShut {NoStop}%
\bibitem [{\citenamefont {Jeong}\ \emph {et~al.}(2016)\citenamefont {Jeong},
  \citenamefont {Ferrante}, \citenamefont {Faleev}, \citenamefont {Samant},
  \citenamefont {Felser},\ and\ \citenamefont {Parkin}}]{Jeong2016}%
  \BibitemOpen
  \bibfield  {author} {\bibinfo {author} {\bibfnamefont {J.}~\bibnamefont
  {Jeong}}, \bibinfo {author} {\bibfnamefont {Y.}~\bibnamefont {Ferrante}},
  \bibinfo {author} {\bibfnamefont {S.~V.}\ \bibnamefont {Faleev}}, \bibinfo
  {author} {\bibfnamefont {M.~G.}\ \bibnamefont {Samant}}, \bibinfo {author}
  {\bibfnamefont {C.}~\bibnamefont {Felser}}, \ and\ \bibinfo {author}
  {\bibfnamefont {S.~S.~P.}\ \bibnamefont {Parkin}},\ }\href {\doibase
  10.1038/ncomms10276} {\bibfield  {journal} {\bibinfo  {journal} {Nat.
  Commun.}\ }\textbf {\bibinfo {volume} {7}},\ \bibinfo {pages} {10276}
  (\bibinfo {year} {2016})}\BibitemShut {NoStop}%
\bibitem [{\citenamefont {Kubota}\ \emph {et~al.}(2019)\citenamefont {Kubota},
  \citenamefont {Kota}, \citenamefont {Ito}, \citenamefont {Umetsu},
  \citenamefont {Sun}, \citenamefont {Mizuguchi},\ and\ \citenamefont
  {Takanashi}}]{Kubota2019}%
  \BibitemOpen
  \bibfield  {author} {\bibinfo {author} {\bibfnamefont {T.}~\bibnamefont
  {Kubota}}, \bibinfo {author} {\bibfnamefont {Y.}~\bibnamefont {Kota}},
  \bibinfo {author} {\bibfnamefont {K.}~\bibnamefont {Ito}}, \bibinfo {author}
  {\bibfnamefont {R.~Y.}\ \bibnamefont {Umetsu}}, \bibinfo {author}
  {\bibfnamefont {M.}~\bibnamefont {Sun}}, \bibinfo {author} {\bibfnamefont
  {M.}~\bibnamefont {Mizuguchi}}, \ and\ \bibinfo {author} {\bibfnamefont
  {K.}~\bibnamefont {Takanashi}},\ }\href {\doibase 10.7567/1882-0786/ab3e2d}
  {\bibfield  {journal} {\bibinfo  {journal} {Appl. Phys. Express}\ }\textbf
  {\bibinfo {volume} {12}},\ \bibinfo {pages} {103002} (\bibinfo {year}
  {2019})}\BibitemShut {NoStop}%
\bibitem [{\citenamefont {Shibata}\ \emph {et~al.}(1973)\citenamefont
  {Shibata}, \citenamefont {Watanabe}, \citenamefont {Yamauchi},\ and\
  \citenamefont {Shinohara}}]{Shibata1973}%
  \BibitemOpen
  \bibfield  {author} {\bibinfo {author} {\bibfnamefont {K.}~\bibnamefont
  {Shibata}}, \bibinfo {author} {\bibfnamefont {H.}~\bibnamefont {Watanabe}},
  \bibinfo {author} {\bibfnamefont {H.}~\bibnamefont {Yamauchi}}, \ and\
  \bibinfo {author} {\bibfnamefont {T.}~\bibnamefont {Shinohara}},\ }\href
  {\doibase 10.1143/jpsj.35.448} {\bibfield  {journal} {\bibinfo  {journal} {J.
  Phys. Soc. Jpn.}\ }\textbf {\bibinfo {volume} {35}},\ \bibinfo {pages} {448}
  (\bibinfo {year} {1973})}\BibitemShut {NoStop}%
\bibitem [{\citenamefont {Umetsu}\ \emph {et~al.}(2014)\citenamefont {Umetsu},
  \citenamefont {Mitsui}, \citenamefont {Yuito}, \citenamefont {Takeuchi},\
  and\ \citenamefont {Kawarada}}]{Umetsu2014}%
  \BibitemOpen
  \bibfield  {author} {\bibinfo {author} {\bibfnamefont {R.~Y.}\ \bibnamefont
  {Umetsu}}, \bibinfo {author} {\bibfnamefont {Y.}~\bibnamefont {Mitsui}},
  \bibinfo {author} {\bibfnamefont {I.}~\bibnamefont {Yuito}}, \bibinfo
  {author} {\bibfnamefont {T.}~\bibnamefont {Takeuchi}}, \ and\ \bibinfo
  {author} {\bibfnamefont {H.}~\bibnamefont {Kawarada}},\ }\href {\doibase
  10.1109/tmag.2014.2321592} {\bibfield  {journal} {\bibinfo  {journal} {{IEEE}
  Trans. Magn.}\ }\textbf {\bibinfo {volume} {50}},\ \bibinfo {pages} {1001904}
  (\bibinfo {year} {2014})}\BibitemShut {NoStop}%
\bibitem [{\citenamefont {Mizukami}\ \emph
  {et~al.}(2013{\natexlab{b}})\citenamefont {Mizukami}, \citenamefont {Sakuma},
  \citenamefont {Kubota}, \citenamefont {Kondo}, \citenamefont {Sugihara},\
  and\ \citenamefont {Miyazaki}}]{Mizukami2013a}%
  \BibitemOpen
  \bibfield  {author} {\bibinfo {author} {\bibfnamefont {S.}~\bibnamefont
  {Mizukami}}, \bibinfo {author} {\bibfnamefont {A.}~\bibnamefont {Sakuma}},
  \bibinfo {author} {\bibfnamefont {T.}~\bibnamefont {Kubota}}, \bibinfo
  {author} {\bibfnamefont {Y.}~\bibnamefont {Kondo}}, \bibinfo {author}
  {\bibfnamefont {A.}~\bibnamefont {Sugihara}}, \ and\ \bibinfo {author}
  {\bibfnamefont {T.}~\bibnamefont {Miyazaki}},\ }\href {\doibase
  10.1063/1.4824031} {\bibfield  {journal} {\bibinfo  {journal} {Appl. Phys.
  Lett.}\ }\textbf {\bibinfo {volume} {103}},\ \bibinfo {pages} {142405}
  (\bibinfo {year} {2013}{\natexlab{b}})}\BibitemShut {NoStop}%
\bibitem [{\citenamefont {Kubota}\ \emph {et~al.}(2020)\citenamefont {Kubota},
  \citenamefont {Kota}, \citenamefont {Ito}, \citenamefont {Umetsu},
  \citenamefont {Sun}, \citenamefont {Mizuguchi},\ and\ \citenamefont
  {Takanashi}}]{Kubota2020}%
  \BibitemOpen
  \bibfield  {author} {\bibinfo {author} {\bibfnamefont {T.}~\bibnamefont
  {Kubota}}, \bibinfo {author} {\bibfnamefont {Y.}~\bibnamefont {Kota}},
  \bibinfo {author} {\bibfnamefont {K.}~\bibnamefont {Ito}}, \bibinfo {author}
  {\bibfnamefont {R.~Y.}\ \bibnamefont {Umetsu}}, \bibinfo {author}
  {\bibfnamefont {M.}~\bibnamefont {Sun}}, \bibinfo {author} {\bibfnamefont
  {M.}~\bibnamefont {Mizuguchi}}, \ and\ \bibinfo {author} {\bibfnamefont
  {K.}~\bibnamefont {Takanashi}},\ }\href {\doibase 10.1063/1.5130388}
  {\bibfield  {journal} {\bibinfo  {journal} {{AIP} {A}dv.}\ }\textbf {\bibinfo
  {volume} {10}},\ \bibinfo {pages} {015122} (\bibinfo {year}
  {2020})}\BibitemShut {NoStop}%
\bibitem [{\citenamefont {Umetsu}\ \emph {et~al.}(2021)\citenamefont {Umetsu},
  \citenamefont {Semboshi}, \citenamefont {Mitsui}, \citenamefont {Katsui},
  \citenamefont {Nozaki}, \citenamefont {Yuitoo}, \citenamefont {Takeuchi},
  \citenamefont {Saito},\ and\ \citenamefont {Kawarada}}]{Umetsu2021a}%
  \BibitemOpen
  \bibfield  {author} {\bibinfo {author} {\bibfnamefont {R.~Y.}\ \bibnamefont
  {Umetsu}}, \bibinfo {author} {\bibfnamefont {S.}~\bibnamefont {Semboshi}},
  \bibinfo {author} {\bibfnamefont {Y.}~\bibnamefont {Mitsui}}, \bibinfo
  {author} {\bibfnamefont {H.}~\bibnamefont {Katsui}}, \bibinfo {author}
  {\bibfnamefont {Y.}~\bibnamefont {Nozaki}}, \bibinfo {author} {\bibfnamefont
  {I.}~\bibnamefont {Yuitoo}}, \bibinfo {author} {\bibfnamefont
  {T.}~\bibnamefont {Takeuchi}}, \bibinfo {author} {\bibfnamefont
  {M.}~\bibnamefont {Saito}}, \ and\ \bibinfo {author} {\bibfnamefont
  {H.}~\bibnamefont {Kawarada}},\ }\href {\doibase
  10.2320/matertrans.mt-m2020309} {\bibfield  {journal} {\bibinfo  {journal}
  {Mater. Trans.}\ }\textbf {\bibinfo {volume} {62}},\ \bibinfo {pages} {680}
  (\bibinfo {year} {2021})}\BibitemShut {NoStop}%
\bibitem [{\citenamefont {Goodenough}\ \emph {et~al.}(1975)\citenamefont
  {Goodenough}, \citenamefont {Street}, \citenamefont {Lee},\ and\
  \citenamefont {Suits}}]{Goodenough1975}%
  \BibitemOpen
  \bibfield  {author} {\bibinfo {author} {\bibfnamefont {J.}~\bibnamefont
  {Goodenough}}, \bibinfo {author} {\bibfnamefont {G.}~\bibnamefont {Street}},
  \bibinfo {author} {\bibfnamefont {K.}~\bibnamefont {Lee}}, \ and\ \bibinfo
  {author} {\bibfnamefont {J.}~\bibnamefont {Suits}},\ }\href {\doibase
  10.1016/0022-3697(75)90073-6} {\bibfield  {journal} {\bibinfo  {journal} {J.
  Phys. Chem. Solids}\ }\textbf {\bibinfo {volume} {36}},\ \bibinfo {pages}
  {451} (\bibinfo {year} {1975})}\BibitemShut {NoStop}%
\bibitem [{\citenamefont {Sun}\ \emph {et~al.}(2020)\citenamefont {Sun},
  \citenamefont {Kubota}, \citenamefont {Ito}, \citenamefont {Takahashi},
  \citenamefont {Hirayama}, \citenamefont {Sonobe},\ and\ \citenamefont
  {Takanashi}}]{Sun2020}%
  \BibitemOpen
  \bibfield  {author} {\bibinfo {author} {\bibfnamefont {M.}~\bibnamefont
  {Sun}}, \bibinfo {author} {\bibfnamefont {T.}~\bibnamefont {Kubota}},
  \bibinfo {author} {\bibfnamefont {K.}~\bibnamefont {Ito}}, \bibinfo {author}
  {\bibfnamefont {S.}~\bibnamefont {Takahashi}}, \bibinfo {author}
  {\bibfnamefont {Y.}~\bibnamefont {Hirayama}}, \bibinfo {author}
  {\bibfnamefont {Y.}~\bibnamefont {Sonobe}}, \ and\ \bibinfo {author}
  {\bibfnamefont {K.}~\bibnamefont {Takanashi}},\ }\href {\doibase
  10.1063/1.5140398} {\bibfield  {journal} {\bibinfo  {journal} {Appl. Phys.
  Lett.}\ }\textbf {\bibinfo {volume} {116}},\ \bibinfo {pages} {062402}
  (\bibinfo {year} {2020})}\BibitemShut {NoStop}%
\bibitem [{\citenamefont {Motizuki}\ \emph {et~al.}(2010)\citenamefont
  {Motizuki}, \citenamefont {Ido}, \citenamefont {Itoh},\ and\ \citenamefont
  {Morifuji}}]{Motizuki2010}%
  \BibitemOpen
  \bibfield  {author} {\bibinfo {author} {\bibfnamefont {K.}~\bibnamefont
  {Motizuki}}, \bibinfo {author} {\bibfnamefont {H.}~\bibnamefont {Ido}},
  \bibinfo {author} {\bibfnamefont {T.}~\bibnamefont {Itoh}}, \ and\ \bibinfo
  {author} {\bibfnamefont {M.}~\bibnamefont {Morifuji}},\ }\href {\doibase
  10.1007/978-3-642-03420-6} {\emph {\bibinfo {title} {Electronic Structure and
  Magnetism of 3d-Transition Metal Pnictides}}}\ (\bibinfo  {publisher}
  {Springer Berlin Heidelberg},\ \bibinfo {year} {2010})\BibitemShut {NoStop}%
\bibitem [{\citenamefont {Ikeda}\ \emph {et~al.}(2010)\citenamefont {Ikeda},
  \citenamefont {Miura}, \citenamefont {Yamamoto}, \citenamefont {Mizunuma},
  \citenamefont {Gan}, \citenamefont {Endo}, \citenamefont {Kanai},
  \citenamefont {Hayakawa}, \citenamefont {Matsukura},\ and\ \citenamefont
  {Ohno}}]{Ikeda2010}%
  \BibitemOpen
  \bibfield  {author} {\bibinfo {author} {\bibfnamefont {S.}~\bibnamefont
  {Ikeda}}, \bibinfo {author} {\bibfnamefont {K.}~\bibnamefont {Miura}},
  \bibinfo {author} {\bibfnamefont {H.}~\bibnamefont {Yamamoto}}, \bibinfo
  {author} {\bibfnamefont {K.}~\bibnamefont {Mizunuma}}, \bibinfo {author}
  {\bibfnamefont {H.~D.}\ \bibnamefont {Gan}}, \bibinfo {author} {\bibfnamefont
  {M.}~\bibnamefont {Endo}}, \bibinfo {author} {\bibfnamefont {S.}~\bibnamefont
  {Kanai}}, \bibinfo {author} {\bibfnamefont {J.}~\bibnamefont {Hayakawa}},
  \bibinfo {author} {\bibfnamefont {F.}~\bibnamefont {Matsukura}}, \ and\
  \bibinfo {author} {\bibfnamefont {H.}~\bibnamefont {Ohno}},\ }\href {\doibase
  10.1038/nmat2804} {\bibfield  {journal} {\bibinfo  {journal} {Nat. Mater.}\
  }\textbf {\bibinfo {volume} {9}},\ \bibinfo {pages} {721} (\bibinfo {year}
  {2010})}\BibitemShut {NoStop}%
\bibitem [{\citenamefont {Sawatzky}\ and\ \citenamefont
  {Street}(1973)}]{Sawatzky1973}%
  \BibitemOpen
  \bibfield  {author} {\bibinfo {author} {\bibfnamefont {E.}~\bibnamefont
  {Sawatzky}}\ and\ \bibinfo {author} {\bibfnamefont {G.~B.}\ \bibnamefont
  {Street}},\ }\href {\doibase 10.1063/1.1662449} {\bibfield  {journal}
  {\bibinfo  {journal} {J. Appl. Phys.}\ }\textbf {\bibinfo {volume} {44}},\
  \bibinfo {pages} {1789} (\bibinfo {year} {1973})}\BibitemShut {NoStop}%
\bibitem [{\citenamefont {Sherwood}\ \emph {et~al.}(1971)\citenamefont
  {Sherwood}, \citenamefont {Nesbitt}, \citenamefont {Wernick}, \citenamefont
  {Bacon}, \citenamefont {Kurtzig},\ and\ \citenamefont
  {Wolfe}}]{Sherwood1971}%
  \BibitemOpen
  \bibfield  {author} {\bibinfo {author} {\bibfnamefont {R.~C.}\ \bibnamefont
  {Sherwood}}, \bibinfo {author} {\bibfnamefont {E.~A.}\ \bibnamefont
  {Nesbitt}}, \bibinfo {author} {\bibfnamefont {J.~H.}\ \bibnamefont
  {Wernick}}, \bibinfo {author} {\bibfnamefont {D.~D.}\ \bibnamefont {Bacon}},
  \bibinfo {author} {\bibfnamefont {A.~J.}\ \bibnamefont {Kurtzig}}, \ and\
  \bibinfo {author} {\bibfnamefont {R.}~\bibnamefont {Wolfe}},\ }\href
  {\doibase 10.1063/1.1660401} {\bibfield  {journal} {\bibinfo  {journal} {J.
  Appl. Phys.}\ }\textbf {\bibinfo {volume} {42}},\ \bibinfo {pages} {1704}
  (\bibinfo {year} {1971})}\BibitemShut {NoStop}%
\bibitem [{\citenamefont {Lee}\ and\ \citenamefont {Suits}(1973)}]{Lee1973}%
  \BibitemOpen
  \bibfield  {author} {\bibinfo {author} {\bibfnamefont {K.}~\bibnamefont
  {Lee}}\ and\ \bibinfo {author} {\bibfnamefont {J.~C.}\ \bibnamefont
  {Suits}},\ }\bibfield  {booktitle} {\emph {\bibinfo {booktitle} {{AIP}
  Conference Proceedings}},\ }\href {\doibase 10.1063/1.2946813} {\bibfield
  {journal} {\bibinfo  {journal} {AIP Conf. Proc.}\ }\textbf {\bibinfo {volume}
  {10}},\ \bibinfo {pages} {1429} (\bibinfo {year} {1973})}\BibitemShut
  {NoStop}%
\bibitem [{\citenamefont {Sun}(2020)}]{Sun2020t}%
  \BibitemOpen
  \bibfield  {author} {\bibinfo {author} {\bibfnamefont {M.}~\bibnamefont
  {Sun}},\ }\emph {\bibinfo {title} {Magnetic properties of {C38}-type
  {Mn}-based intermetallic compound films for spintronic applications}},\
  \href@noop {} {\bibinfo {type} {{PhD} thesis}},\ \bibinfo  {school} {Tohoku
  University} (\bibinfo {year} {2020})\BibitemShut {NoStop}%
\bibitem [{\citenamefont {Yuasa}\ and\ \citenamefont
  {Djayaprawira}(2007)}]{Yuasa2007}%
  \BibitemOpen
  \bibfield  {author} {\bibinfo {author} {\bibfnamefont {S.}~\bibnamefont
  {Yuasa}}\ and\ \bibinfo {author} {\bibfnamefont {D.~D.}\ \bibnamefont
  {Djayaprawira}},\ }\href {\doibase 10.1088/0022-3727/40/21/r01} {\bibfield
  {journal} {\bibinfo  {journal} {J. Phys. D: Appl. Phys.}\ }\textbf {\bibinfo
  {volume} {40}},\ \bibinfo {pages} {R337} (\bibinfo {year}
  {2007})}\BibitemShut {NoStop}%
\bibitem [{\citenamefont {Yamada}, \citenamefont {Kubota},\ and\ \citenamefont
  {Nakatani}(2020)}]{Yamada2020}%
  \BibitemOpen
  \bibfield  {author} {\bibinfo {author} {\bibfnamefont {K.}~\bibnamefont
  {Yamada}}, \bibinfo {author} {\bibfnamefont {K.}~\bibnamefont {Kubota}}, \
  and\ \bibinfo {author} {\bibfnamefont {Y.}~\bibnamefont {Nakatani}},\ }\href
  {\doibase 10.1063/5.0005472} {\bibfield  {journal} {\bibinfo  {journal} {J.
  Appl. Phys.}\ }\textbf {\bibinfo {volume} {127}},\ \bibinfo {pages} {133906}
  (\bibinfo {year} {2020})}\BibitemShut {NoStop}%
\bibitem [{\citenamefont {Kubota}\ \emph {et~al.}(2021)\citenamefont {Kubota},
  \citenamefont {Ito}, \citenamefont {Umetsu}, \citenamefont {Mizuguchi},\ and\
  \citenamefont {Takanashi}}]{Kubota2021}%
  \BibitemOpen
  \bibfield  {author} {\bibinfo {author} {\bibfnamefont {T.}~\bibnamefont
  {Kubota}}, \bibinfo {author} {\bibfnamefont {K.}~\bibnamefont {Ito}},
  \bibinfo {author} {\bibfnamefont {R.~Y.}\ \bibnamefont {Umetsu}}, \bibinfo
  {author} {\bibfnamefont {M.}~\bibnamefont {Mizuguchi}}, \ and\ \bibinfo
  {author} {\bibfnamefont {K.}~\bibnamefont {Takanashi}},\ }\href {\doibase
  10.1063/9.0000138} {\bibfield  {journal} {\bibinfo  {journal} {{AIP} {A}dv.}\
  }\textbf {\bibinfo {volume} {11}},\ \bibinfo {pages} {015124} (\bibinfo
  {year} {2021})}\BibitemShut {NoStop}%
\bibitem [{\citenamefont {Tsunekawa}\ \emph {et~al.}(2005)\citenamefont
  {Tsunekawa}, \citenamefont {Djayaprawira}, \citenamefont {Nagai},
  \citenamefont {Maehara}, \citenamefont {Yamagata}, \citenamefont {Watanabe},
  \citenamefont {Yuasa}, \citenamefont {Suzuki},\ and\ \citenamefont
  {Ando}}]{Tsunekawa2005}%
  \BibitemOpen
  \bibfield  {author} {\bibinfo {author} {\bibfnamefont {K.}~\bibnamefont
  {Tsunekawa}}, \bibinfo {author} {\bibfnamefont {D.~D.}\ \bibnamefont
  {Djayaprawira}}, \bibinfo {author} {\bibfnamefont {M.}~\bibnamefont {Nagai}},
  \bibinfo {author} {\bibfnamefont {H.}~\bibnamefont {Maehara}}, \bibinfo
  {author} {\bibfnamefont {S.}~\bibnamefont {Yamagata}}, \bibinfo {author}
  {\bibfnamefont {N.}~\bibnamefont {Watanabe}}, \bibinfo {author}
  {\bibfnamefont {S.}~\bibnamefont {Yuasa}}, \bibinfo {author} {\bibfnamefont
  {Y.}~\bibnamefont {Suzuki}}, \ and\ \bibinfo {author} {\bibfnamefont
  {K.}~\bibnamefont {Ando}},\ }\href {\doibase 10.1063/1.2012525} {\bibfield
  {journal} {\bibinfo  {journal} {Appl. Phys. Lett.}\ }\textbf {\bibinfo
  {volume} {87}},\ \bibinfo {pages} {072503} (\bibinfo {year}
  {2005})}\BibitemShut {NoStop}%
\bibitem [{\citenamefont {Sakuraba}\ \emph {et~al.}(2007)\citenamefont
  {Sakuraba}, \citenamefont {Hattori}, \citenamefont {Oogane}, \citenamefont
  {Kubota}, \citenamefont {Ando}, \citenamefont {Sakuma}, \citenamefont
  {Telling}, \citenamefont {Keatley}, \citenamefont {van~der Laan},
  \citenamefont {Arenholz}, \citenamefont {Hicken},\ and\ \citenamefont
  {Miyazaki}}]{Sakuraba2007}%
  \BibitemOpen
  \bibfield  {author} {\bibinfo {author} {\bibfnamefont {Y.}~\bibnamefont
  {Sakuraba}}, \bibinfo {author} {\bibfnamefont {M.}~\bibnamefont {Hattori}},
  \bibinfo {author} {\bibfnamefont {M.}~\bibnamefont {Oogane}}, \bibinfo
  {author} {\bibfnamefont {H.}~\bibnamefont {Kubota}}, \bibinfo {author}
  {\bibfnamefont {Y.}~\bibnamefont {Ando}}, \bibinfo {author} {\bibfnamefont
  {A.}~\bibnamefont {Sakuma}}, \bibinfo {author} {\bibfnamefont {N.~D.}\
  \bibnamefont {Telling}}, \bibinfo {author} {\bibfnamefont {P.}~\bibnamefont
  {Keatley}}, \bibinfo {author} {\bibfnamefont {G.}~\bibnamefont {van~der
  Laan}}, \bibinfo {author} {\bibfnamefont {E.}~\bibnamefont {Arenholz}},
  \bibinfo {author} {\bibfnamefont {R.~J.}\ \bibnamefont {Hicken}}, \ and\
  \bibinfo {author} {\bibfnamefont {T.}~\bibnamefont {Miyazaki}},\ }\href
  {\doibase 10.3379/jmsjmag.31.338} {\bibfield  {journal} {\bibinfo  {journal}
  {J. Magn. Soc. Jpn.}\ }\textbf {\bibinfo {volume} {31}},\ \bibinfo {pages}
  {338} (\bibinfo {year} {2007})}\BibitemShut {NoStop}%
\bibitem [{\citenamefont {Kubota}\ \emph {et~al.}(2011)\citenamefont {Kubota},
  \citenamefont {Mizukami}, \citenamefont {Watanabe}, \citenamefont {Wu},
  \citenamefont {Zhang}, \citenamefont {Naganuma}, \citenamefont {Oogane},
  \citenamefont {Ando},\ and\ \citenamefont {Miyazaki}}]{Kubota2011mg}%
  \BibitemOpen
  \bibfield  {author} {\bibinfo {author} {\bibfnamefont {T.}~\bibnamefont
  {Kubota}}, \bibinfo {author} {\bibfnamefont {S.}~\bibnamefont {Mizukami}},
  \bibinfo {author} {\bibfnamefont {D.}~\bibnamefont {Watanabe}}, \bibinfo
  {author} {\bibfnamefont {F.}~\bibnamefont {Wu}}, \bibinfo {author}
  {\bibfnamefont {X.}~\bibnamefont {Zhang}}, \bibinfo {author} {\bibfnamefont
  {H.}~\bibnamefont {Naganuma}}, \bibinfo {author} {\bibfnamefont
  {M.}~\bibnamefont {Oogane}}, \bibinfo {author} {\bibfnamefont
  {Y.}~\bibnamefont {Ando}}, \ and\ \bibinfo {author} {\bibfnamefont
  {T.}~\bibnamefont {Miyazaki}},\ }\href {\doibase 10.1063/1.3603034}
  {\bibfield  {journal} {\bibinfo  {journal} {J. Appl. Phys.}\ }\textbf
  {\bibinfo {volume} {110}},\ \bibinfo {pages} {013915} (\bibinfo {year}
  {2011})}\BibitemShut {NoStop}%
\bibitem [{\citenamefont {Wen}\ \emph {et~al.}(2016)\citenamefont {Wen},
  \citenamefont {Hadorn}, \citenamefont {Okabayashi}, \citenamefont {Sukegawa},
  \citenamefont {Ohkubo}, \citenamefont {Inomata}, \citenamefont {Mitani},\
  and\ \citenamefont {Hono}}]{Wen2016CFA}%
  \BibitemOpen
  \bibfield  {author} {\bibinfo {author} {\bibfnamefont {Z.}~\bibnamefont
  {Wen}}, \bibinfo {author} {\bibfnamefont {J.~P.}\ \bibnamefont {Hadorn}},
  \bibinfo {author} {\bibfnamefont {J.}~\bibnamefont {Okabayashi}}, \bibinfo
  {author} {\bibfnamefont {H.}~\bibnamefont {Sukegawa}}, \bibinfo {author}
  {\bibfnamefont {T.}~\bibnamefont {Ohkubo}}, \bibinfo {author} {\bibfnamefont
  {K.}~\bibnamefont {Inomata}}, \bibinfo {author} {\bibfnamefont
  {S.}~\bibnamefont {Mitani}}, \ and\ \bibinfo {author} {\bibfnamefont
  {K.}~\bibnamefont {Hono}},\ }\href {\doibase 10.7567/apex.10.013003}
  {\bibfield  {journal} {\bibinfo  {journal} {Appl. Phys. Express}\ }\textbf
  {\bibinfo {volume} {10}},\ \bibinfo {pages} {013003} (\bibinfo {year}
  {2016})}\BibitemShut {NoStop}%
\bibitem [{\citenamefont {Husain}\ \emph {et~al.}(2019)\citenamefont {Husain},
  \citenamefont {Sisodia}, \citenamefont {Chaurasiya}, \citenamefont {Kumar},
  \citenamefont {Singh}, \citenamefont {Yadav}, \citenamefont {Akansel},
  \citenamefont {Chae}, \citenamefont {Barman}, \citenamefont {Muduli},
  \citenamefont {Svedlindh},\ and\ \citenamefont {Chaudhary}}]{Husain2019}%
  \BibitemOpen
  \bibfield  {author} {\bibinfo {author} {\bibfnamefont {S.}~\bibnamefont
  {Husain}}, \bibinfo {author} {\bibfnamefont {N.}~\bibnamefont {Sisodia}},
  \bibinfo {author} {\bibfnamefont {A.~K.}\ \bibnamefont {Chaurasiya}},
  \bibinfo {author} {\bibfnamefont {A.}~\bibnamefont {Kumar}}, \bibinfo
  {author} {\bibfnamefont {J.~P.}\ \bibnamefont {Singh}}, \bibinfo {author}
  {\bibfnamefont {B.~S.}\ \bibnamefont {Yadav}}, \bibinfo {author}
  {\bibfnamefont {S.}~\bibnamefont {Akansel}}, \bibinfo {author} {\bibfnamefont
  {K.~H.}\ \bibnamefont {Chae}}, \bibinfo {author} {\bibfnamefont
  {A.}~\bibnamefont {Barman}}, \bibinfo {author} {\bibfnamefont {P.~K.}\
  \bibnamefont {Muduli}}, \bibinfo {author} {\bibfnamefont {P.}~\bibnamefont
  {Svedlindh}}, \ and\ \bibinfo {author} {\bibfnamefont {S.}~\bibnamefont
  {Chaudhary}},\ }\href {\doibase 10.1038/s41598-018-35832-3} {\bibfield
  {journal} {\bibinfo  {journal} {Sci. Rep.}\ }\textbf {\bibinfo {volume}
  {9}},\ \bibinfo {pages} {1085} (\bibinfo {year} {2019})}\BibitemShut
  {NoStop}%
\bibitem [{\citenamefont {Ellingham}(1944)}]{Ellingham1944}%
  \BibitemOpen
  \bibfield  {author} {\bibinfo {author} {\bibfnamefont {H.~J.~T.}\
  \bibnamefont {Ellingham}},\ }\href {\doibase 10.1002/jctb.5000630501}
  {\bibfield  {journal} {\bibinfo  {journal} {J. Soc. Chem. Ind.}\ }\textbf
  {\bibinfo {volume} {63}},\ \bibinfo {pages} {125} (\bibinfo {year}
  {1944})}\BibitemShut {NoStop}%
\bibitem [{\citenamefont {Seki}\ \emph {et~al.}(2010)\citenamefont {Seki},
  \citenamefont {Kubota}, \citenamefont {Fukushima}, \citenamefont {Yakushiji},
  \citenamefont {Yuasa}, \citenamefont {Ando}, \citenamefont {Maehara},
  \citenamefont {Yamagata}, \citenamefont {Okuyama},\ and\ \citenamefont
  {Tsunekawa}}]{Seki2010}%
  \BibitemOpen
  \bibfield  {author} {\bibinfo {author} {\bibfnamefont {T.}~\bibnamefont
  {Seki}}, \bibinfo {author} {\bibfnamefont {H.}~\bibnamefont {Kubota}},
  \bibinfo {author} {\bibfnamefont {A.}~\bibnamefont {Fukushima}}, \bibinfo
  {author} {\bibfnamefont {K.}~\bibnamefont {Yakushiji}}, \bibinfo {author}
  {\bibfnamefont {S.}~\bibnamefont {Yuasa}}, \bibinfo {author} {\bibfnamefont
  {K.}~\bibnamefont {Ando}}, \bibinfo {author} {\bibfnamefont {H.}~\bibnamefont
  {Maehara}}, \bibinfo {author} {\bibfnamefont {S.}~\bibnamefont {Yamagata}},
  \bibinfo {author} {\bibfnamefont {H.}~\bibnamefont {Okuyama}}, \ and\
  \bibinfo {author} {\bibfnamefont {K.}~\bibnamefont {Tsunekawa}},\ }\href
  {\doibase 10.1063/1.3524543} {\bibfield  {journal} {\bibinfo  {journal} {J.
  Appl. Phys.}\ }\textbf {\bibinfo {volume} {108}},\ \bibinfo {pages} {123915}
  (\bibinfo {year} {2010})}\BibitemShut {NoStop}%
\end{thebibliography}%

\end{document}